\documentclass[twocolumn,english,aps,superscriptaddress, pra,groupedaddress]{revtex4-1}

\usepackage[latin9]{inputenc}
\usepackage{amsmath}
\usepackage{amssymb}
\usepackage{graphicx}
\usepackage{babel}
\usepackage{mathrsfs}
\usepackage{amsfonts}
\usepackage{epstopdf}
\usepackage{caption}
\usepackage{multirow}

\begin{document}
\title{Far-Away-From-Equilibrium Quantum Critical  Conformal Dynamics:\\
Reversibility, Thermalization, and Hydrodynamics}

\author{Jeff Maki$^{1, 2)}$}
\author{Fei Zhou$^{2)}$}
\affiliation{1) Department of Physics and HKU-UCAS Joint Institute for Theoretical and Computational Physics, The University of Hong Kong, Hong Kong, China  \\ 2) Department of Physics and Astronomy, University of British Columbia, 6224 Agricultural Road, Vancouver, BC, V6T 1Z1, Canada}

\begin{abstract}
Generic far-away-from-equilibrium many-body dynamics involve entropy production, and hence are thermodynamically irreversible. Near quantum critical points, an emergent conformal symmetry can impose strong constraints on entropy production rates, and in some cases completely forbid entropy production, which usually occurs for systems that deviate from quantum critical points. In this article, we illustrate how the vanishing entropy production near a quantum critical point results in reversible far-away-from-equilibrium dynamics at finite temperatures that are otherwise irreversible. Away from the quantum critical point, the quantum dynamics are damped, and our analysis directly relates the thermalization time scale to the hydrodynamic viscosity near quantum critical points with dynamical critical exponent $z=2$. We demonstrate how both controllable reversible and irreversible dynamics can be potentially studied in cold gas experiments using Feshbach resonances.



\end{abstract}

\date{\today}
\maketitle

\section{Introduction}

In a generic many-body system, a net entropy production usually occurs in far-away-from-equilibrium dynamics, leading to irreversible processes. A well known example is the free expansion of a weakly interacting gas inside a box, when the gas is originally confined in, say, the right half of the box. This energy conserved process is irreversible since the total entropy increases  by $N \ln 2$, where $N (\rightarrow \infty)$  is the number of particles. The entropy production simply reflects the exponentially larger number of micro-states at a given energy associated with the whole box compared to half of the box; there are $~2^N$ times more micro-states in the whole box. The exponential increase of micro-states results in an exponentially small probability for observing all $N$ particles in either half of the box at any time, and hence the dynamics are irreversible.

It is possible to have a similar general paradigm for quantum dynamics. Let us consider a generic interacting Fermi gas that is in thermal equilibrium within a portion of the box, and let it expand into the rest of the box. Such a state is highly excited from the point of view of a thermal state of $N$-fermions in the whole box. The excitation energy per particle will then be finite, and the resulting dynamics will be far-away-from-equilibrium. During the expansion, the initial state can be projected onto the $N$-Fermion states in the whole box with the same average total energy, $E$. At such a large energy, the density of states, $D(E, N)$, is exponentially large. Indeed, a direct count shows that for energies $E/N \gg \epsilon_F$, the density of states, $D(E, N \rightarrow \infty )$,  scales as:   

\begin{equation}
D(E, N) \sim \frac{1}{E} \exp(\frac{3}{2} N \ln (E/\epsilon_F) )  
\end{equation}

\noindent for a three dimensional box. Here $\epsilon_F$ is the Fermi energy of an $N$-particle gas in the whole box of volume $V$, and $E$ is an extensive quantity. 

In the standard thermalization paradigm, the initial state will explore the canonical ensemble with average total energy, $E$. The interactions, weak or strong, generically lead to an effective thermal mixture of these micro-states after a sufficiently long time. The physical properties measured after the thermalization time scale typically do not depend on the specifics of the initial state (except the initial energy), and are thus robust. Such quantum dynamics are irreversible for the same reason as the classical gas; there is an exponentially large number of micro-states for energy, $E$. The probability to observe the gas in its initial state is then practically zero, and the dynamics are irreversible. It is then natural to assert that the dynamics resulted in an increase in entropy due to the larger number of micro-states. 


\begin{figure}
\includegraphics[scale=0.38]{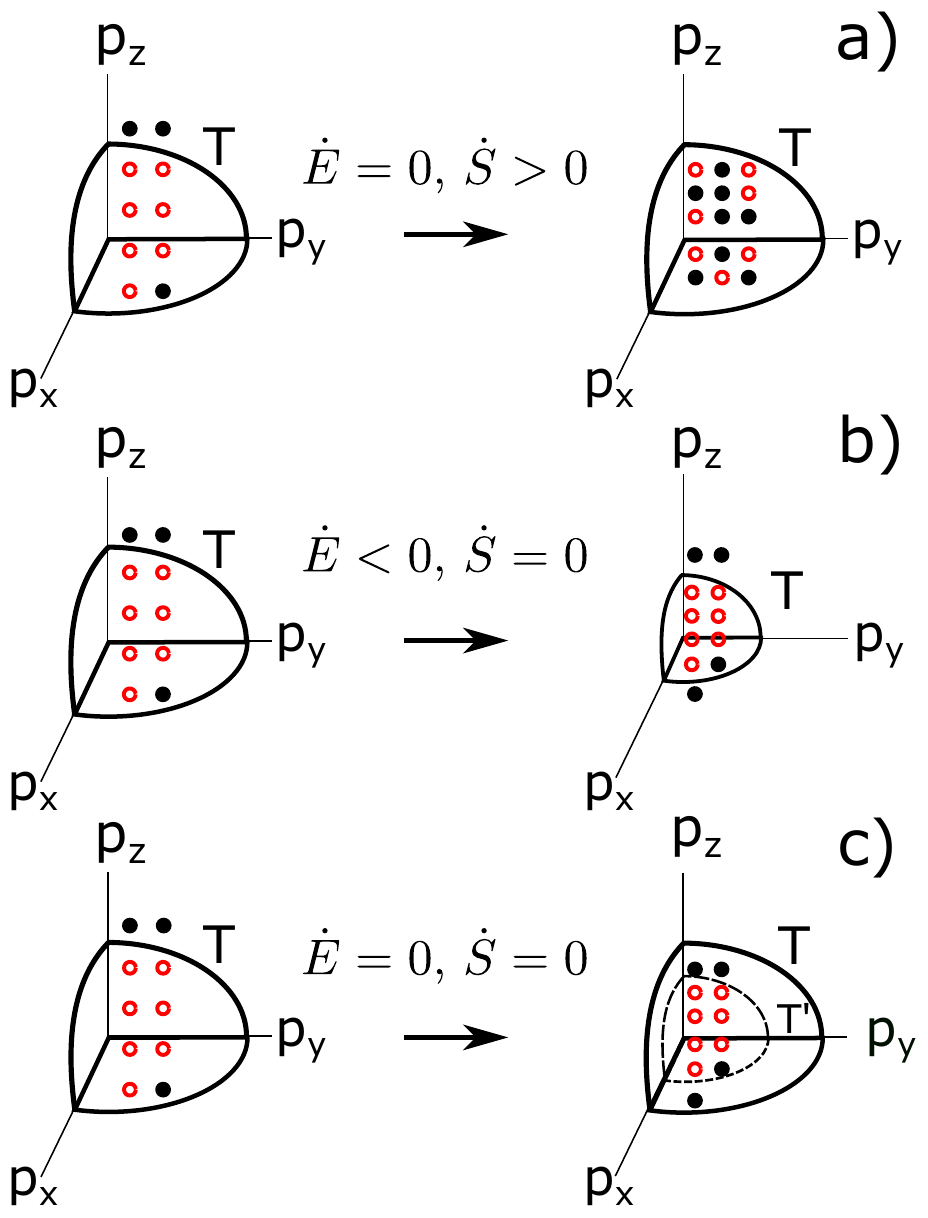}
\caption{Illustration of entropy dynamics during a) generic free expansion, b) an adiabatic expansion of weakly interacting fermions, and c) in a free conformal expansion ({\it fCE}) near a critical point. The energy, $E$, is conserved in both a) and c) but not in b). Each point in a)-c) represents a one-partcle state that is either occupied (red open circle) or unoccupied (solid dot). The number of possible $N$-body micro-states is $e^S$ where $S$ is the entropy. The thick solid line indicates the temperature, $T$, far beyond which states are rarely occupied. In a), the number of allowed micro-states states is exponentially larger after expansion. In b), the energy is reduced because of work done, but the number of occupied $N$-body micro-states is constant. In c), the Boltzmann temperature, $T$, is rescaled to $T'$, in a way identical to b) so that the entropy is conserved; however, the energy is also conserved since no work has been done just as in a) (see also Ref.~\cite{Energy}).}
\label{fig:schematic}
\end{figure}

In this article, we will focus on similar far-away-from-equilibrium quantum dynamics for a gas near a quantum critical point. We will illustrate that unlike the free expansion examples discussed above, releasing such a quantum critical gas into the vacuum can occur with {\em zero} entropy production. Such a peculiar {\it free} expansion which conserves both the energy and entropy is obviously distinct from the previously discussed free expansion, which always comes with entropy production, see Fig.~(\ref{fig:schematic}) a). It is also distinct from typical adiabatic expansion processes in thermodynamics which conserve entropy, but {\em do not} conserve energy, Fig.~(\ref{fig:schematic}) b). To distinguish this peculiar energy and entropy preserving expansion from the free and adiabatic expansions defined in the standard thermodynamic context \cite{Landau_thermo}, we name such an expansion {\em free conformal expansion} ({\it fCE}), Fig.~(\ref{fig:schematic}) c).


Our main conclusions are two-fold. First, when the interactions are tuned to a quantum critical point with dynamical exponent $z=2$, a {\it fCE} can be fully reversible under a time reversible action. Second, away from the quantum critical point, a hydrodynamic analysis suggests irreversible dynamics where thermalization occurs at a time scale, $1/\Gamma$, that is inversely proportional to the bulk viscosity, $\zeta$, i.e. $1/\Gamma  \sim 1/\zeta$. This time scale approaches infinity when approaching a conformal symmetric critical point. 

Although our conclusions apply to a number of strongly interacting scale invariant quantum systems, we will primarily be focused on the application to the resonant Fermi gas in three dimensions. At resonance and in the zero-range limit, the Fermi gas is well described by a quantum critical point with dynamical exponent $z=2$ \cite{Sachdev}. This should be differentiated from the traditional unitary Fermi gas that does not have the exact conformal symmetry if the gas is near but not exactly at resonance. It is beneficial to study this system as it can be precisely manipulated by a number of experimental techniques \cite{Bloch08,Chin10}. For this reason, both the reversible dynamics at the critical point and the irreversible dynamics away from the critical point can be readily studied by using Fermi gases and Feshbach resonances.

The remainder of the article is organized as follows. Sec.~\ref{sec:prod} discusses how the conformal and $SO(2,1)$ symmetries equate the expansion dynamics to the dynamics of a gas governed by a fictitious projective Hamiltonian. We then discuss how this relation is equivalent to entropy conservation and {\it fCE}. In Sec.~\ref{sec:rev} we examine the prospect of reversible dynamics for harmonically trapped unitary Fermi gases. In Sec.~\ref{sec:irr} we then consider how the breaking of scale invariance leads to irreversible dynamics for harmonically trapped Fermi gases. Our conclusions are presented in Sec.~\ref{sec:conc}.

\section{Entropy Production in Conformal Expansion}
\label{sec:prod}

The quantum dynamics that we will discuss are dictated by emergent non-relativistic scale and conformal symmetries \cite{Hagen72, Niederer72, Henkel94} near a critical point with dynamical critical exponent $z=2$. Many-body systems related to such a quantum critical point can be strongly interacting electrons near Lifshitz transitions driven by external pressures \cite{Lifshitz60}, or scale invariant quantum gases, both in the presence and absence of harmonic trapping potentials. Some primary examples of such quantum gases are three-dimensional atomic gases with short-ranged s-wave interactions at a Feshbach resonance \cite{Chin10, Sachdev}, or strongly interacting quantum gases in one dimension \cite{Minguzzi05, Kheruntsyan,TG19}.

Generally, quantum critical points with emergent Galilean invariance exhibit $SO(2,1)$ conformal symmetry \cite{Rosch97, Castin04, Castin06, Nishida07, Gritsev10}, in addition to the usual scale symmetries. One remarkable consequence of conformal symmetry is the absence of the bulk viscosity in the hydrodynamics of systems in their normal state  \cite{Wingate06,Son07,Taylor12}, although the shear viscosity remains finite \cite{Thomas11, Enss11, Thomas14}. Recently, the vanishing bulk viscosity has been further explicitly related to the emergence of strongly interacting conformal tower states, and consequential density matrix dynamics  \cite{Maki19}. Indeed, another distinct feature of the general SO(2,1) conformal dynamics is that the density matrix will maintain the same amount of information during expansion. A conformal quantum fluid can therefore undergo a free expansion into the vacuum with both energy and entropy conserved; in stark contrast with the free expansion of a classical thermal gas into a vacuum, where the entropy always increases.
 
To facilitate this discussion, we consider a resonantly interacting Fermi gas initially confined in a three dimensional harmonic trap of frequency, $\omega_0$, and at thermal equilibrium with temperature, $T_0$. Since the gas is in thermal equilibrium, the initial density matrix is a mixed state that is governed by the Boltzmann distribution. In the position representation the initial $N$-body density matrix can be written as:

\begin{eqnarray}
\rho_{eq} \left(\left\lbrace {\bf r}_i \right\rbrace, \left\lbrace {\bf r}'_i \right\rbrace, i=1...N; \frac{\omega_0}{T_0} \right), \nonumber \\
\label{DM0}
\end{eqnarray}

\noindent where the subscript denotes the density matrix is of the equilibrium form. The Hamiltonian describing the system initially is given by: $H_s + \omega_0^2 C$, where we define the scale invariant Hamiltonian for the system, $H_s$, in the absence of the harmonic trapping potential, $C$. In our case, the scale invariant Hamiltonian, $H_s$, and trapping potential, $C$, are given by:

\begin{align}
H_s &= \int d{\bf r} \ \psi^{\dagger}({\bf r}) \left( - \frac{\nabla^2}{2} \right) \psi({\bf r}) \nonumber \\
&+ \int d{\bf r} d{\bf r'} \psi^{\dagger}({\bf r})\psi^{\dagger}({\bf r'}) V_s({\bf r - r'}) \psi({\bf r'})\psi({\bf r}) \nonumber \\
C &= \int d{\bf r} \ \psi^{\dagger}({\bf r}) \frac{r^2}{2} \psi({\bf r})
\label{eq:H}
\end{align}

\noindent where $\psi^{(\dagger)}({\bf r})$ is  the second quantized annihilation (creation) operator, and $V_s({\bf r-r'})$ is a scale invariant two-body potential. An example of such a potential is a short ranged isotropic interaction with infinite s-wave scattering length, $a_s = \infty$. We have also suppressed the spin indices as they do not play a major role in our discussions.

\begin{figure}
\includegraphics[scale=0.58]{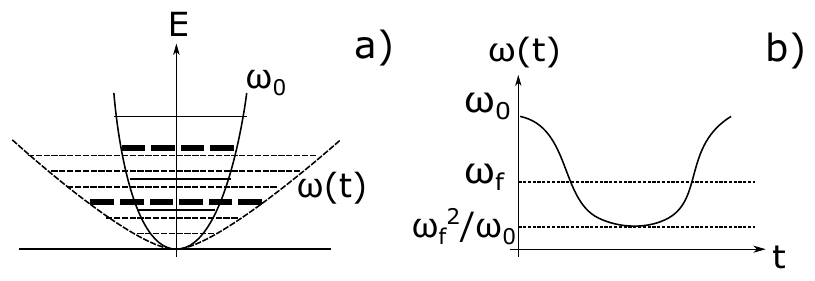}
\caption{a) The many-body spectra during the conformal expansion.
Solid lines are  for the initial potential, and dashed lines are for the projective Hamiltonian with $\omega(t)$ in Eq.~(\ref{proj}). The (dashed) bold thick line indicates the temperature, ($T(t)$) $T_0$. b) $\omega(t)$ of the projective Hamiltonian after a quench of the trapping potential from $\omega_0$ to $\omega_f$. The conformal dynamics follows this projective trap frequency $\omega(t)$ while the physical trap frequency, $\omega_f$, is a constant shown as the dashed line.}
\label{fig:entropy}
\end{figure}

A consequence of the SO(2,1) conformal symmetry is that once the gas has been released into {\em free} space without a confining potential, the physical state of the gas at an arbitrary time, $t$, can be mapped onto an equilibrium state of a fictitious projective Hamiltonian, $H_{proj}(t)$, at temperature, $T(t)$ \cite{Maki19}. The mapping is exact up to a position dependent gauge factor that induces a hydrodynamic current \cite{Castin06}. The projective Hamiltonian is given by:

\begin{eqnarray}
H_{proj}(t) =H_s + \frac{1}{2} \omega^2(t) C
\label{proj}
\end{eqnarray}

\noindent where $H_s$ and $C$ are both defined in Eq.~(\ref{eq:H}). The physical Hamiltonian, $H_{phys}$, under which the dynamics are studied, is simply $H_s$, i.e. $H_{phys}=H_{proj}(\omega(t)=0)$. In this representation both the frequency, $\omega(t)$, and the temperature, $T(t)$, are rescaled in a time dependent fashion:

\begin{align}
\hbar \omega (t) &=  \frac{\hbar \omega_0}{\lambda^2(t)} & 
T(t) &=\frac{T_0}{\lambda^2(t)}
\label{FT}
\end{align}

\noindent where $\lambda(t)$ is a dynamical rescaling factor that is given by $(1 + \omega_0^2 t^2)^{1/2}$ for the expansion into free space. The projective equilibrium temperature, $T(t)$ \cite{Energy}, is rescaled in an identical fashion to $\omega(t)$. Although the states of the projective Hamiltonian become denser in energy space as $\omega(t)$ decreases, the Boltzmann weight for a given state is constant since $\omega(t)/T(t)$ is invariant in time. This is depicted in Fig.~(\ref{fig:entropy}). Therefore, the $N$-particle density matrix at time $t$ takes a simple form:

\begin{eqnarray}
&& \rho_N\left(\left\lbrace {\bf r}_i \right\rbrace, \left\lbrace {\bf r'}_i \right\rbrace, i=1...N; t\right) \nonumber \\
&&=G\frac{1}{\lambda^{3N}(t)} {\rho}_{eq} \left(\left\lbrace \frac{{\bf r}_i}{\lambda(t)} \right\rbrace, \left\lbrace \frac{{\bf r}'_i}{\lambda(t)} \right\rbrace, i=1...N; \frac{\omega (t)}{T(t)} \right) \nonumber \\
\label{DM}
\end{eqnarray}

\noindent From Eq.~(\ref{DM}), one can see that the time-dependent density matrix is obtained from its initial equilibrium value, Eq.~(\ref{DM0}), via a time-dependent rescaling of the position coordinates, and a gauge transformation, $G$, that is irrelevant to our discussions.

The conformal structure of Eq.~(\ref{DM}) suggests that entropy must be conserved, as the information content in the density matrix is unchanged. To see this more clearly we note that the entropy, $S(t)$, is defined via:

\begin{equation}
S(t)= \int d{\bf r} S({\bf r},t)  = - Tr\left[\rho_N \cdot \ln \rho_N \right]
\label{eq:entropy}
\end{equation}

\noindent where $Tr$ denotes the trace over the $N$ position coordinates, and $\ln \rho_N$ is the log of the $N$-body density matrix. In Eq.~(\ref{eq:entropy}) we have also defined an entropy density, $S({\bf r},t)$, which can be obtained by performing a partial trace over $N-1$ coordinates in Eq.~(\ref{eq:entropy}). It is straightforward to verify that the dynamics of the density matrix in Eq.~(\ref{DM}) translate into similar dynamics for the entropy density:

\begin{equation}
S({\bf r},t) = \frac{1}{\lambda^3(t)} S\left(\frac{{\bf r}}{\lambda(t)},0\right),
\label{eq:entropy_2}
\end{equation}

\noindent which conserves the total entropy. Such a conclusion was obtained in Ref.~\cite{Maki19}, using the SO(2,1) symmetry to obtain a differential equation for the entropy density. Here we provide an alternative description using the projective Hamiltonian as it  provides a clear intuitive picture for the conservation of entropy.

It is important to note that Eq.~(\ref{eq:entropy_2}) has a caveat. Although for the convenience of presentation we have assumed that the gas is initially in thermal equilibrium in an isotropic harmonic potential, this condition is not essential. For example, if the initial state is in thermal equilibrium in an anisotropic harmonic potential, there will be entropy production at earlier stages of the dynamics. However, the rate of entropy production in this case quickly vanishes in the long time limit, resulting in an asymptotic {\it fCE} with an emergent conformal symmetry. In contrast, when the scale symmetry is explicitly broken, there will be a finite entropy production even in the long time limit \cite{Maki19, Maki18}. For the remainder of this work, we will focus on the case of Fermi gases initially in thermal equilibrium inside isotropic harmonic potentials as entropy is conserved exactly for scale invariant interactions.

Below we will explore a specific consequence of entropy conservation; the prospect of reversible dynamics, and how the breaking of scale symmetry undoubtedly leads to irreversible and entropy producing dynamics. The reversible far-away-from equilibrium quantum dynamics we propose below is, to our knowledge, one of very few examples, if not the only one, that is available in strongly interacting quantum many-body systems. 
The scale and conformal symmetry, on the other hand, can also be present in non-interacting quantum systems. In that case, the non-equilibrium quantum dynamics can also be reversible \cite{Maki18}. In classical thermal gases, the possibility of reversible entropy conserving dynamics was pointed out by Ludwig Boltzmann more than a century ago. More recently, such reversible dynamics were observed in classical gases in a beautiful experiment at JILA \cite{Lobser15}. The reversible dynamics which we will discuss next and which are based on conformal symmetry, can be considered a quantum version of the Boltzmann breather in a strongly interacting quantum gas, a {\em quantum Boltzmann breather}.

\section{Reversible Dynamics}
\label{sec:rev}

In order to discuss reversible dynamics, we need a time-reversal symmetric Hamiltonian. To this end we consider a resonantly interacting, three-dimensional, normal Fermi gas, which is equilibrated at temperature $T_0$. The gas is also subject to the following time-dependent harmonic trapping frequency:

\begin{align}
\Omega(t) &= \left \lbrace\begin{array}{lll} \omega_0 & & |t| > t_h/2 \\											  \omega_f & & |t|<t_h/2 \end{array} \right.
\label{Action}
\end{align}


\noindent where  $\omega_0 \gg \omega_f$. A schematic for this trapping potential is shown in Fig.~(\ref{fig:thermalization}) a). The initial trap restricts the fermions in a small spatial region for $t<-t_h/2$. At $t=-t_h/2$, the fermions are released into a much flatter trap with frequency, $\omega_{f}$. After a hold time, $t_h$, the trapping potential reverts to its initial frequency. The benefit of Eq.~(\ref{Action}) is that it mimics releasing the $N$-fermion state from a small region into a much larger region, and allows for the gas to be recollected in the original spatial region at a later time.


The initial state of the system is still given by Eq.~(\ref{DM0}). As shown in Appendix \ref{app:proj}, the dynamics of this many-body system can still be projected onto an equilibrium state defined by a {\em  fictitious projective} Hamiltonian, $H_{proj}(t)$, with a projective trapping frequency, $\omega(t)$, given by Eq.~(\ref{FT}) - just like the expansion into the vacuum. For this case $\lambda(t)$ is given by:

\begin{align}
\lambda(t)=\left[ \cos^2 \left(\omega_f (t+\frac{t_h}{2})\right) +\frac{\omega_0^2}{\omega_F^2} \sin^2 \left(\omega_f (t+\frac{t_h}{2}) \right)\right]^{1/2} \nonumber \\
\label{lambda}
\end{align}

\noindent for $|t|<t_h/2$. As stated previously, the statistical temperature, $T(t)$, will have the same time dependence as $\omega(t)$. 

Although the physical state of the system can be mapped onto an equilibrium state of the projective Hamiltonian up to a gauge, generally $H_{proj}(t)$ is distinctly different from the physical Hamiltonian $H_{phys}  = H_{proj}(\omega(t) = \omega_f)$ between $-t_h/2$ and $t_h/2$ \cite{current}, as shown in Fig.~(\ref{fig:entropy}) b). Hence the dynamic states are highly excited from the point of view of $H_{phys}$, and are truly far-away-from-equilibrium. 

To highlight this point, consider an initial state with exponentially small entropy, or $T_0 \ll \hbar \omega_0$. The dynamic state at $ t+t_h/2=\pi/(2 \omega_f) $ will be dilated by a factor: $\lambda(t)=\omega_0/\omega_f$, corresponding to an equilibrium state of the {\em fictitious} potential: $\omega(t)=\omega_f^2/\omega_0$. However, this state is $\left(\omega_0/\omega_f\right)^{1/2} (\gg 1)$ larger than the size of the ground state of the physical trap, $\omega_f$. Furthermore, as the entropy remains nearly zero during the expansion, this dynamic state can not be related to an equilibrium state of $H_{phys}$, with the same larger size, as it would have a much higher entropy. This concludes that the dynamic state is truly far-away-from-equilibrium.


Let us now examine the motion of a Fermi gas in the trapping potential given by Eq.~(\ref{Action}). We summarize the results of the reversible dynamics in Fig.~(\ref{fig:thermalization}).  For $|t| < t_h/2$, the dynamics of the density matrix is given by Eq.~(\ref{DM}) with the time-dependent rescaling factor defined in Eq.~(\ref{lambda}). The gas will exhibit undamped oscillations at a frequency $2\omega_f$.

These perfect oscillations go beyond the breathing and collective mode physics previously discussed in the literature \cite{Bartenstein04, Kinast04, Grimm07, Grimm13b, Stringari04, Giorgini08, Kinast05, Hu04, Odelin99, Pedri03, Landau}. To understand the difference, we first note that Eqs.~(\ref{DM}) and (\ref{lambda}) describe highly non-linear, far-away-from-equilibrium dynamics, as we are interested in the limit of $\omega_f \ll \omega_0$. This is in contrast to standard collective mode physics which are usually explored around equilibrium via linearized equations of motion \cite{Odelin99,Landau,Pedri03}. 

However, if one considers the limit when $\omega_f \lesssim \omega_0$, and linearizes the conformal dynamics in Eqs.~(\ref{DM}) and (\ref{lambda}), the conformal dynamics predict an {\em undamped} breathing mode with a frequency {\em pinned} at $2\omega_0$. This undamped breathing mode has been both predicted and observed for scale invariant quantum gases \cite{Rosch97,Castin04, Castin06, Grimm07, Bartenstein04, Kinast04, Lobser15}. Here we stress that the dynamics at resonance are protected by conformal symmetry, and always exhibit the forementioned undamped oscillations at $2\omega_0$. Our results are hence valid in both the hydrodynamic and collisionless regime, defined below, as long as the interactions are scale invariant. Another case where dynamics are constrained by conformal symmetry is for nearly non-interacting gases. Indeed, it is well-known that a classical thermal gas can support a undamped Boltzmann breather which had been recently observed in cold gases \cite{Lobser15}.

On the other hand, for generic strongly interacting gases away from resonance, the breathing mode frequency is {\em no longer pinned} at $2\omega_0$, and in fact can substantially deviate from $2 \omega_0$ \cite{Pedri03, Odelin99}. Such deviation near resonance will also be addressed in the hydrodynamic limit discussed below. In the same context, we also find these collective modes in general will have finite damping in the absence of conformal symmetry, a consequence of non-vanishing entropy production. 

Generically, the collective mode damping rate, $\Gamma$, is a function of the relaxation time $\tau$ \cite{Landau, Nozieres}. At high temperatures, $T\gg \epsilon_F$ where $\epsilon_F$ is the Fermi energy, one can estimate  $\tau^{-1} \sim n/T^{1/2}$  while at temperatures $T\ll \epsilon_F$, $\tau^{-1} \sim T^2/n^{2/3}$. In the hydrodynamic limit, $\omega_0 \tau \to 0$, the damping rate is $\Gamma \propto \omega_0^2 \tau$, while in the collisionless limit, $\omega_0 \tau \to \infty$, $\Gamma \propto 1/\tau$. In Table~\ref{tab:1} we compare the conformal solution to the dynamics, Eq.~(\ref{lambda}), and the collective mode physics.

\begin{table}[]
\begin{tabular}{|c|c|c|c|}
\hline
              & Frequency    & $\Gamma$ & Linear  \\ \hline
Hydrodynamic  &  $\neq 2 \omega_0$            &  Finite                      & Yes     \\ \hline
Collisionless & $\neq 2 \omega_0$            &   Finite                     & Yes     \\ \hline
Conformal     & $=2 \omega_0$ &    $Zero$                    & No    \\ \hline
\end{tabular}
\caption{Comparison of the breathing mode in the hydrodynamic and collisionless regimes in a generic interacting quantum gas away from resonance, and the symmetry-protected conformal dynamics at resonance. We compare the frequency, the damping rate $\Gamma$, and whether the motion is described by a linearized equation of motion. Finite damping away from resonance in collective modes is indicated by the finite entropy production in dynamics.}
\label{tab:1}
\end{table}

For times $t > t_h/2$, there are two possibilities for the long-time dynamics. If the holding time, $t_h$, is matched to $n\pi/\omega_f$ for some integer $n$, the density matrix, Eq.~(\ref{DM}), will exactly return to its initial thermal equilibrium form, right as the trap returns to its initial value. In this case, the dynamics not only conserve energy and entropy, but it is possible to completely retrieve the initial quantum state. This retrieval is a full many-body effect related to the conformal symmetry. For other values of $t_h$, $\lambda(t)$ will further oscillate at a frequency $\omega_0$ after the initial trap is restored, as the gas will not be in an equilibrium state right after the second quench. These oscillations will also be undamped as the dynamics still conserve entropy.

Eqs.~(\ref{DM}) and (\ref{lambda}) are indicative of the existence of $N$-body conformal tower states studied before \cite{Castin04, Castin06, Nishida07, Maki18,Maki19}. The SO(2,1) symmetry suggests that the spectrum of $H_{proj}(\omega(t) = \omega_f)$ can be divided into a number of towers, each labeled by conserved quantities such as the angular momentum etc. Each tower hosts a set of states that are equally spaced with a universal spacing $2 \omega_f$, independent of the label for an individual tower. Following Eqs.~(\ref{DM}) and (\ref{lambda}), the density matrix dynamics only involves  frequencies of $2 n \omega_f$ for $n = 0, \pm 1, \pm 2...$. The many-body dynamics with a single fundamental frequency can only occur if the eigenstates of $H_{prof}(\omega(t) = \omega_f)$  have this equally spaced character (i.e. the states are equally spaced within each tower labelled by a conserved angular momentum).


\section{Irreversible Dynamics and Thermalization Time}
\label{sec:irr}
To contrast the above reversible far-away-from-equilibrium dynamics due to the emergent conformal symmetry with more conventional irreversible dynamics, we further examine
the dynamics of away-from-resonance Fermi gases in the normal phase when subjected to a quantum quench in the trapping frequency: $\omega_0$ to $\omega_f \ll \omega_0$ at $t=0$. 
In this case, the conformal symmetry is explicitly broken by a finite correlation length, $\xi$, or equivalently by the large but finite scattering length, $a_s$. We will focus on the hydrodynamic limit where the dynamics at frequency $\omega_f$  is much slower than the intrinsic scattering rate, $\omega_f \tau \ll 1$, which is valid for a wide range of temperatures near resonance, where atoms scatter most frequently. In this case, the entropy is no longer conserved resulting in a finite bulk viscosity, $\zeta$. The hydrodynamic viscosity near resonance has been carefully calculated in a number of thorough studies \cite{Schaefer13, Nishida19, Enss19, Hofmann20} which we refer to for explicit details. The results obtained there are consistent with a general analysis based on the breaking of conformal symmetry near quantum critical points \cite{Maki19}. 

Here we examine the dynamics of the scaling parameter in the presence of entropy production. Applying the standard hydrodynamic techniques \cite{Thomas14, Stringari02,O'Hara02} to the moment of inertia, $\langle r^2 \rangle(t) = \lambda^2(t) \langle r^2 \rangle(0)$, one can obtain the following differential equation for the scaling parameter:

\begin{align}
\frac{d^2 \lambda^2(t)}{dt^2} &= 2\left[\omega_0^2 + \omega_f^2\right] -4 \omega_f^2 \lambda^2(t) \nonumber \\
&+ \Delta\tilde{P} \left(\frac{1}{\lambda(t)}-1\right)- \tilde{\zeta} \frac{d \lambda^2(t)}{dt};
\label{eq:hydrodynamics}
\end{align}

\noindent where $\Delta \tilde{P} = 6 \int d^3r \Delta P(0)/\left(m N \langle r^2 \rangle(0)\right)$ is due to the deviation of the pressure from its scale invariant value, $\tilde{\zeta} = 9\int d^3r \zeta(r,0)/\left(m N \langle r^2 \rangle(0)\right)$ is the bulk viscosity, and $m$ is the single particle mass. For more details see Appendix \ref{app:hydro}. Eq.~(\ref{eq:hydrodynamics}) is the main result on hydrodynamics near a conformal symmetric, strongly interacting, critical point.

At the critical point, when $\Delta \tilde{P} = 0$ and $\tilde{\zeta} = 0$, the hydrodynamic solution for $\lambda^2(t)$ is identical to Eq.~(\ref{lambda}), the full quantum solution suggested by conformal symmetry. Therefore, the hydrodynamic solution is fully consistent with the conformal structure of the density matrix in Eq.~(\ref{DM}), and the microscopic conformal tower states discussed previously \cite{Castin04, Castin06, Nishida07, Maki18, Maki19}. 

It is worth remarking that the shear stress tensor does not enter the above equation for the scaling parameter $\lambda^2(t)$ and makes no contributions to dynamics here. Thus, even if the initial state is not perfectly isotropic (and as a result there can be entropy production, and dissipation in other degrees of freedom or higher moments due to the finite shear viscosity), the dynamics associated with $\lambda^2 (t)$ are still reversible as the bulk viscosity in Eq.~(\ref{eq:hydrodynamics}) vanishes exactly when the interactions are tuned to a scale invariant critical point, such as Feshbach resonance. Moreover, one can further show this is also generally true even in the collisionless limit, i.e. the frequency of oscillation and the damping dynamics associated with $\lambda^2(t)$ are constrained by conformal symmetry when the interactions are at a scale invariant critical point. The absence of the damping in the moment of inertia, and the independence of this result on the initial conditions, can also be seen generally from the Heisenberg equation of motion, where the SO(2,1) conformal symmetry gaurantees both the frequency and persistence of the motion. For more details see Appendix~\ref{app:HEOM}.

Away from the resonant critical point, the bulk viscosity becomes finite; following \cite{Landau}, the entropy production can be expressed as:

\begin{equation}
\frac{\partial S(t)}{\partial t} = \frac{9}{4T_0} \int d^3r \zeta(r,0) \left( \frac{d\lambda^2(t)}{dt} \right)^2.
\label{eq:hydro_entropy}
\end{equation}


\begin{figure}
\includegraphics[scale=0.54]{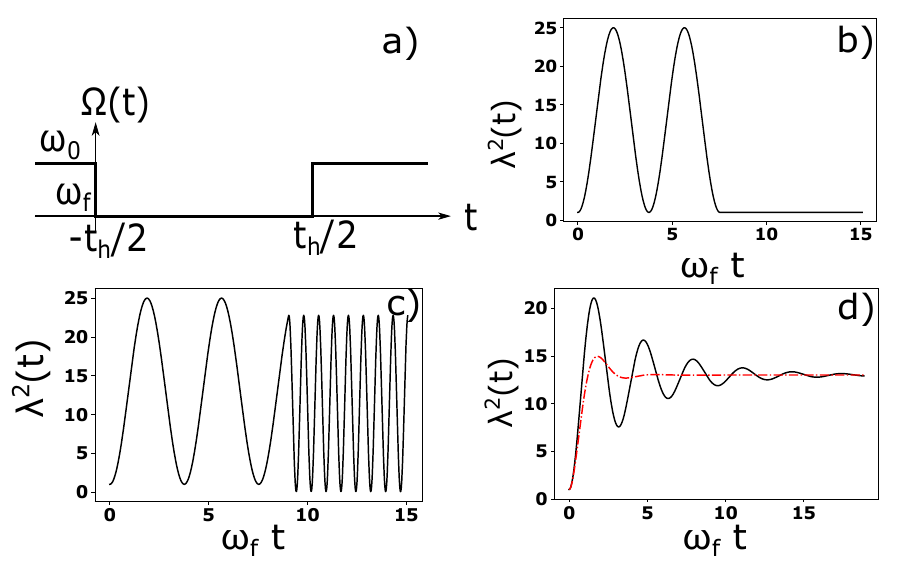}
\caption{Reversible versus irreversible non-equilibrium dynamics. a) The frequency of a harmonic potential in a time-reversible cycle, Eq.~(\ref{Action}). b), c) the scaling factors for reversible dynamics as a function of time when $t_h$ is commensurate and incommensurate with $\pi/\omega_f$, respectively. We have chosen $\omega_0 = 5\omega_f$ for the simulations. d) The scaling factor for reversible dynamics for viscosities $\tilde{\zeta} = 0.1$ (solid black line) and $\tilde{\zeta} = 0.4$ (red dashed line) when $t_h \rightarrow \infty$. The thermalization time, $\tau$, is inversely proportional to the viscosity coefficient $\tilde{\zeta}$. }
\label{fig:thermalization}
\end{figure}

Therefore, the inclusion of the bulk viscosity, $\tilde{\zeta}$, in the hydrodynamic equation, allows one to explicitly examine irreversible dynamics in the vicinity of critical points, and to investigate the relation between the thermalization time scale and entropy production, or equivalently the viscosity. The results are shown in Fig.~(\ref{fig:thermalization}) d) where we have solved Eq.~(\ref{eq:hydrodynamics}) numerically. 

There are several general features of the dynamics away from resonance. First, the gas oscillates at a frequency:

\begin{align}
\omega &\approx 2\omega_f \left(1 + \beta\right), &  
\beta &= \frac{\Delta \tilde{P}}{4\omega_f^2}
\end{align}

\noindent which is valid at $O(1/a_s)$. Here $\beta \sim 1/a_s$ measures the deviation from the conformal symmetric point, or resonance. As one can see, the frequency of the oscillations is no longer pinned at $2 \omega_f$ due to the broken conformal symmetry. Secondly, at long times the scaling parameter always approaches a finite value: 

\begin{equation}
\lambda^2(t=\infty) \approx \frac{1}{2}\left(\frac{\omega_0^2}{\omega_f^2} +1\right) +\beta\left(\sqrt{\frac{2\omega_f^2}{\omega_0^2+\omega_f^2}}-1\right),
\label{eq:asymp_lambda}
\end{equation}

\noindent which is also correct to $O(1/a_s)$, and is independent of the entropy production rate and the viscosity, $\zeta$. Therefore, in hydrodynamics it is natural to conclude that the gas has thermalized in the final harmonic potential with frequency $\omega_f$.

The damping of the oscillations is exponential.  As suggested explicitly by Eq.~(\ref{eq:hydrodynamics}), the thermalization time inferred from this analysis scales as:

\begin{eqnarray}
\frac{1}{\Gamma} \approx \frac{2}{\tilde{\zeta}} \propto a_s^2 \gg \tau
\label{eq:tau}
\end{eqnarray}

\noindent where the second scaling relation is also directly inferred by the dynamical critical exponent $z=2$. Although $1/\Gamma$ becomes infinite as $a_s$ becomes infinite at critical points, the intrinsic Boltzmannian relaxation time, $\tau$, is always finite and short. Again we stress this result is only valid near resonance when $a_s \to \infty$. Therefore, the thermalization dynamics observed here offers a rather convenient and direct way to measure hydrodynamic coefficients, and vice versa.

\begin{figure}
\includegraphics[scale=0.55]{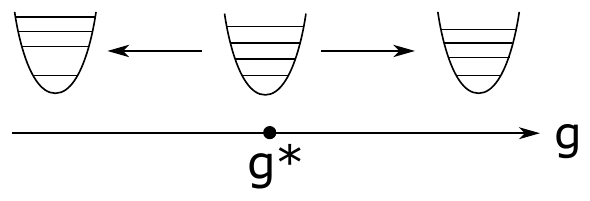}
\caption{Schematic of conformal tower states. Exactly at the critical point $1/a_s=0$ with an emergent $SO(2,1)$ symmetry, the spectrum consists of a set of equally spaced conformal tower states; we only show one of the towers explicitly. For large but finite values of $a_s$, the tower states are randomly scrambled (See main text). Dashed lines indicate the quantum critical regime in the density (n) - coupling constant (g) plane. We note that a similar picture occurs near the non-interacting scale invariant point, $a_s$ = 0.}
\label{fig:qcp}
\end{figure}

\section{Conclusions}
\label{sec:conc}

The thermalization suggested by the hydrodynamic solution at finite scattering lengths implies that many-body chaos has likely developed among the highly excited states. Even in a perfect harmonic trap, the number of highly excited micro-states with energy $E/N \gg N^{1/3}$ is exponentially large: $D_H (E, N) \sim \exp( 3N \ln (E/N^{1/3}) )$, much like a box potential. Away from the quantum critical point, interactions in this highly-excited manifold are expected to scramble the conformal tower states within each tower, see Fig.~(\ref{fig:qcp}), and likely result in a band of states with exponentially small energy differences. This will result in the thermalization of the quantum gas. We expect the physical properties will not depend on the initial quantum state, nor will the initial state be retrievable at any large time \cite{recurrence}.

 
Indeed, in a previous study \cite{Maki18} we have shown that a microscopic description of the dynamics in the vicinity of critical points can be fully characterized by a universal $V$ matrix, or a four-fermion interaction operator that is a natural extension of the thermodynamic contact \cite{Tan08, Zhang09}. Following the general random matrix approach to highly excited states, we speculate that the $V$-matrix will have a similar character. Namely, the highly excited spectrum within a given conformal tower will be scrambled according to random ensemble theories \cite{Mehta04}. This random scrambling could result in an exponentially large number of randomly shifted eigenfrequencies. We have indeed numerically observed Wigner-Dyson statistics induced by a random four-fermion operator, or V-matrix, acting on 70 many-body conformal tower states. 

Such many-body chaos can lead to an effective thermalization in an isolated quantum system. What has been observed in the hydrodynamic analysis is fully consistent with this general paradigm. In this regard, the emergent conformal symmetry appears to be sufficient to prevent far-away-from-equilibrium states from being thermalized, although the system is not integrable.

F.Z. was in part supported by the Canadian Institute for Advanced Research. We would also like to thank Gora Shlyapnikov and Shizhong Zhang for useful discussions. We also thank Shizhong Zhang for supporting J. M. through the HK GRF 17318316, 17305218, CRF C6026-16W and C6005-17G, and the Croucher Foundation under the Croucher Innovation Award.

\appendix

\numberwithin{equation}{section}
\renewcommand\theequation{\Alph{section}.\arabic{equation}}

\section{Derivation of the Projective Hamiltonian}
\label{app:proj}

In this appendix we explicitly show that the dynamics of harmonically confined quantum gases subjected to a quantum quench of the trapping potential follow a projective Hamiltonian. Consider a gas that is initially confined in a harmonic potential with frequency, $\omega_0$. At $t=0$, the potential is quenched to a much shallower trap with frequency, $\omega_f \ll \omega_0$. 

For gases that are initially in thermal equilibrium, the initial density matrix is diagonal in the energy eigenstates of the initial Hamiltonian. In order to study the post quench dynamics it is sufficient to consider the dynamics of the individual eigenstates. Let us define:

\begin{align}
\left[H_s + \omega^2_0 C \right] \left| n, \omega_0 \right\rangle &= E^0_n  \left| n, \omega_0 \right\rangle
\label{def}
\end{align}

\noindent where $H_s$ and $C$ are the scale invariant Hamiltonian, and the harmonic trapping potential, respectively (see below for formal definitions). We also define $|n, \omega_0 \rangle$ as an eigenstate of $H_s +\omega_0^2 C$, with energy, $E_n$, and $n$ is a collective index for all the quantum numbers. We now consider the motion of this state under the post quench Hamiltonian:

\begin{align}
\left| n, \omega_0,t \right\rangle (t) &= e^{-i(H_s + \omega_f^2 C) t} \left| n, \omega_0 \right\rangle.
\label{conformal_t}
\end{align}

\noindent From Eq.~(\ref{def}) and (\ref{conformal_t}) one can show that the time-evolved eigenstate must satisfy:

\begin{align}
E_n^0 \left| n, \omega_0,t \right\rangle =  &\left[e^{-i(H_s + \omega_f^2 C)t} \left(H_s + \omega_0^2 C\right) e^{+i(H_s + \omega_f^2 C)t} \right] \nonumber \\
& \left| n, \omega_0,t \right\rangle \nonumber \\
\label{eq:1}
\end{align}

\noindent In order to evaluate Eq.~(\ref{eq:1}), we follow the approach used in Ref.~\cite{Maki19}; it is necessary to employ the identity:

\begin{equation}
e^A B e^{-A} = B + [A,B] + \frac{1}{2!}[A,[A,B]] +...
\end{equation}

\noindent as well as the SO(2,1) algebra:

\begin{align}
[H_s, C] &= -i D & [D, H_s] &= 2i H_s & [D, C] &= -2i C
\label{algebra} 
\end{align}

\noindent where $H_s$, $C$, and  $D$ are defined as:

\begin{align}
H_s &= \int d{\bf r} \psi^{\dagger}({\bf r}) \left(-\frac{ \nabla^2}{2} \right) \psi({\bf r}) \nonumber \\
&+ \frac{1}{2}\int d{\bf r} d{\bf r'} \ V_s({\bf r-r'}) \psi^{\dagger}({\bf r}) \psi^{\dagger}({\bf r}') \psi^{\dagger}({\bf r'}) \psi^{\dagger}({\bf r}) \nonumber \\
D &=-i\int d{\bf r} \psi^{\dagger}({\bf r}) \left[ {\bf r} \cdot \nabla + \frac{3}{2}\right] \psi({\bf r}) \nonumber \\
C&=\int d{\bf r} \psi^{\dagger}({\bf r}) \frac{r^2}{2} \psi({\bf r}).
\label{eq:def_so21}
\end{align}

\noindent Above we have defined $\psi^{(\dagger)}({\bf r})$ as the second quantized annihilation (creation) operator, and $V_s$, as a  scale invariant two-body potential. We have also suppressed the spin indices for simplicity.

Thanks to the SO(2,1) algebra, it is possible to obtain a closed expression for the operator in Eq.~(\ref{eq:1}):

\begin{align}
e^{i(H_s + \omega_f^2 C)t} & \left(H_s + \omega_0^2 C\right) e^{-i(H_s + \omega_f^2 C)t} = H_s + \omega_0^2 C \nonumber \\
&+ \frac{\omega_0^2-\omega_f^2}{2\omega_f^2} \left(1 - \cos(2 \omega_f t) \right) \left( H_s - \omega_f^2 C\right)  \nonumber \\
& - \frac{\omega_0^2-\omega_f^2}{2\omega_f} \sin(2\omega_f t) D
\label{eq:2}
\end{align}

Let us now define:

\begin{equation}
\lambda(t) = \sqrt{\cos(\omega_f t)^2 + \frac{\omega_0^2}{\omega_f^2} \sin(\omega_f t)^2}
\end{equation}

\noindent Rearranging the terms in Eq.~(\ref{eq:2}) one can obtain:

\begin{align}
e^{i(H_s + \omega_f^2 C)t} & \left(H_s + \omega_0^2 C\right) e^{-i(H_s + \omega_f^2 C)t} \nonumber \\
&= \lambda^2(t) \left[ H_s + \omega_f^2 C\right] -\dot{\lambda}(t) \lambda(t) D \nonumber \\
&+ \left[\ddot{\lambda}(t) \lambda(t) + \dot{\lambda}^2(t)\right] C
\end{align}

\noindent For the discussion of the instantaneous Hamiltonian, we note that Eq.~(\ref{eq:2}) can be cast in the form:

\begin{equation}
\lambda^2(t) \left[\tilde{H}_s + \frac{\omega_0^2}{\lambda^4(t)} C\right]
\label{eq:3}
\end{equation}

\noindent where:

\begin{align}
\tilde{H}_s &= \frac{1}{2}\int d{\bf r} \psi^{\dagger}({\bf r}) \left(-i \nabla - \frac{\dot{\lambda}(t)}{\lambda(t)} {\bf r} \right)^2 \psi({\bf r}) \nonumber \\
&+ \frac{1}{2}\int d{\bf r} d{\bf r'} \ V_s({\bf r-r'}) \psi^{\dagger}({\bf r}) \psi^{\dagger}({\bf r}') \psi^{\dagger}({\bf r'}) \psi^{\dagger}({\bf r})
\end{align}

\noindent is the boosted scale invariant Hamiltonian, and we have used the identity:

\begin{equation}
\ddot{\lambda}(t) \lambda^3(t) + \omega_f^2 \lambda^4(t) = \omega_0^2
\end{equation}

\noindent The effect of the boost is irrelevant to our discussions, as we are only interested in the diagonal components of the $N$-body density matrix. 

Eq.~(\ref{eq:3}) also implies:

\begin{equation}
\frac{E_n^0}{\lambda^2(t)} \left| n, \omega_0,t \right\rangle = \left[\tilde{H}_s + \frac{\omega^2_0}{\lambda^4(t)} C \right] \left| n, \omega_0,t \right\rangle
\label{final}
\end{equation}

\noindent which is equivalent to the projective Hamiltonian shown in Eqs.~(\ref{proj}) and (\ref{FT}), up to a boost. Therefore the eigenstate of the initial trap, will remain an eigenstate of the instantaneous Hamiltonian, up to a position dependent gauge factor.

As mentioned previously, for a quantum thermal gas initially in equilibrium at temperature $T_0$, the density matrix can be written as a canonical ensemble of energy eigenstates with the Boltzmann weighting:

\begin{equation}
\rho(0) = \sum_n e^{-\frac{E_n}{T_0}} |n, \omega_0 \rangle \langle n ,\omega_0 |
\end{equation}

\noindent where Boltzmann's constant has been set to one. Using Eq.~(\ref{final}), one can show that the time evolved density matrix still has the equilibrium form:

\begin{align}
\rho(t) &= \sum_n e^{-E^0_n/T_0} e^{-i(H_s + \omega_f^2 C)t} |n, \omega_0 \rangle \langle n ,\omega_0 | e^{-i(H_s + \omega_f^2 C)t} \nonumber \\
&= \sum_n \exp\left[-\frac{E^0_n}{\lambda^2(t)} \frac{\lambda^2(t)}{T_0}\right] |n, \omega_0,t \rangle \langle n, \omega_0,t|
\end{align}

\noindent which still possesses the equilibrium form, with a time-dependent projective temperature, $T(t)$. Casting this result into the position representation will give the desired result. The density matrix satisfies Eq.~(\ref{DM}) with $\lambda(t)$ given by Eq.~(\ref{lambda}).

\section{Hydrodynamics of the Moment of Inertia}
\label{app:hydro}

In this appendix we use the hydrodynamic equations of motion to obtain Eq.~(\ref{eq:hydrodynamics}). Eq.~(\ref{eq:hydrodynamics}) describes the motion of a nearly unitary Fermi gas after a quantum quench in the trap frequency:

\begin{align}
\Omega(t) &= \left \lbrace\begin{array}{lll} \omega_0 & & t < 0 \\											  \omega_f & & 0<t \end{array} \right.
\end{align}

\noindent 
We begin by defining the hydrodynamic average of the moment of inertia as:

\begin{equation}
\langle r^2 \rangle(t) = \int \frac{d {\bf r}}{N} r^2 n({\bf r},t),
\label{B:moment}
\end{equation}


\noindent where $N$ is the total number of particles in the system, and $n({\bf r}, t)$ is the density. In hydrodynamics, the dynamics of the density, $n({\bf r},t)$ and the velocity field, ${\bf v}$, are determined by the following conservation law and equations of motion \cite{Landau},

\begin{align}
\partial_t n + \partial_{i} \cdot \left(n v_i\right) &= 0 \nonumber \\
m n \left(\partial_t + v_j \cdot \nabla_j \right) v_i &= -\partial_i P - n \Omega^2(t) r_i \nonumber \\
&+\sum_j \partial_j\left(\eta \sigma_{i,j} + \zeta \sigma' \delta_{i,j}\right).
\label{B:hydro_eom}
\end{align}

\noindent where we have suppressed the dependence of the position and time coordinates in all of the thermodynamic quantities. In Eq.~(\ref{B:hydro_eom}) $m$ is the atomic mass, $P$ is the pressure, and $\eta$ and $\zeta$ are the shear and bulk viscosities respectively.  We have also defined the shear stress tensors as
\begin{equation}
\sigma_{i,j} = \frac{1}{2} (\partial_i v_j + \partial_j v_i)-\frac{1}{3} \sigma' \delta_{i,j} , \sigma' = \nabla \cdot {\bf v}. 
\end{equation}

From Eqs.~(\ref{B:moment}) and (\ref{B:hydro_eom}) one can show:

\begin{align}
\frac{m}{2}\frac{d^2 \langle r^2 \rangle(t)}{dt^2} &= 2\left[\frac{3}{2} \int \frac{d{\bf r}}{N} P + \frac{m}{2}\langle v^2 \rangle\right] \nonumber \\
&- \Omega^2(t)\langle r^2\rangle(t) - 3 \int \frac{d{\bf r}}{N} \zeta({\bf r},t) \nabla \cdot {\bf v}
\label{B:moment_2}
\end{align}

\noindent In Eq.~(\ref{B:moment_2}), the hydrodynamic average of the kinetic energy, $\langle v^2 \rangle$, is defined similarly to Eq.~(\ref{B:moment}).
In the limit when $\Omega(t)=0$, this set of equation was previously introduced in Ref.~\cite{Thomas14} to study hydrodynamics of expanding Fermi gases.

In order to simplify Eq.~(\ref{B:moment_2}) we use the fact that energy is conserved for $t>0$:

\begin{align}
0 &= \frac{d}{dt} \langle H\rangle(t) \nonumber  \\
0 &= \frac{d}{dt} \left[ \int \frac{d{\bf r}}{N} \  \mathcal{E}(t) + \frac{m}{2}\langle v^2 \rangle(t)  + \frac{\Omega^2(t)}{2} \langle r^2 \rangle(t)\right]
\end{align}

\noindent where $\mathcal{E}$ is the energy density. Next we relate the energy density to the pressure by noting that for a scale invariant systems: $\mathcal{E} = 3/2 P$. However, the breaking of scale invariance naturally introduces a shift in the pressure, $\Delta P$ such that:

\begin{equation}
\mathcal{E} = \frac{3}{2} (P-\Delta P).
\end{equation}

\noindent The identification of the pressure then lets us  obtain the following result for $t>0$:

\begin{align}
\frac{1}{2}\frac{d^2 \langle r^2 \rangle(t)}{dt^2} &= 2 \langle H \rangle(0^+) - 2\omega_f^2 \langle r^2 \rangle(t) \nonumber \\
&-\int \frac{d{\bf r}}{N} \zeta({\bf r},t) \nabla \cdot {\bf v} + 3 \int \frac{d{\bf r}}{N} \Delta P(t).
\end{align}

It is possible to relate $\langle H \rangle(0^+)$ to the initial conditions. For $t<0$ the system is in thermal equilibrium, i.e. ${\bf v} =0$. After examining Eq.~(\ref{B:moment_2}) for $t<0$, one can show that:

\begin{equation}
2 \langle H \rangle(0^-)  = 2\omega_0^2  \langle r^2 \rangle(0^-) - 3 \int \frac{d{\bf r}}{N} \Delta P(0).
\label{B:H0}
\end{equation}

\noindent Eq.~(\ref{B:H0}) allows one to determine the size and energy of the gas right after the trapping potential is quenched:

\begin{align}
\langle r^2 \rangle(0^{+}) &= \langle r^2 \rangle(0^{-}) \nonumber \\
\langle H \rangle(0^+) &= \langle H \rangle (0^{-})+ \frac{1}{2}\left(\omega_f^2 - \omega_0^2 \right) \langle r^2 \rangle(0).
\label{B:r0}
\end{align}

\noindent Eqs.~(\ref{B:H0}) and (\ref{B:r0}) result in the full description of the moment of inertia for $t>0$:

\begin{align}
m\frac{d\langle r^2 \rangle(t)}{dt^2} &= 2(\omega_0^2 + \omega_f^2) \langle r^2 \rangle(0) - 4 \omega_f^2 \langle r^2 \rangle(t) \nonumber \\
&+6 \int \frac{d{\bf r}}{N} \left[ \Delta P(t) - \Delta P(0)\right] \nonumber \\
&- 6\int \frac{d {\bf r}}{N} \zeta({\bf r},t) \nabla \cdot {\bf v}
\label{B:moment_3}
\end{align}

For nearly scale invariant systems, it natural to assume that the solution of Eq.~(\ref{B:moment_3}) still has a scaling form. Consider the following scaling ansatz

\begin{align}
n({\bf r},t) &= \frac{1}{\lambda^3(t)} n\left(\frac{{\bf r}}{\lambda(t)},0\right) & {\bf v} &= \frac{\dot{\lambda}(t)}{\lambda(t)}{\bf r} & T(t) &= \frac{T_0}{\lambda^2(t)} \nonumber \\
\label{B:scaling_ansatz}
\end{align}

\noindent where $\lambda(0) = 1$, and $\dot{\lambda}(0) = 0$. Using this ansatz one obtains an expression for the time-dependent scaling parameter:

\begin{align}
\frac{d^2 \lambda^2(t)}{dt^2} &=  2(\omega_0^2 + \omega_f^2) - 4\omega_f^2 \lambda^2(t) \nonumber \\
&+ \frac{6}{m\langle r^2 \rangle(0)} \int \frac{d{\bf r}}{N} \left[\Delta P(t) - \Delta P(0) \right] \nonumber \\
&- \frac{9}{m\langle r^2 \rangle(0)} \int \frac{d {\bf r}}{N} \zeta({\bf r},t) \frac{1}{\lambda^2(t)} \frac{d\lambda^2(t)}{dt}
\label{B:eom_1}
\end{align}

\noindent In Eq.~(\ref{B:eom_1}), we note that the shift in the pressure, $\Delta P(t)$, is proportional to $1/a_s$, and the bulk viscosity, $\zeta({\bf r},t)$, is proportional to $1/a_s^2$.

If one were to take the scale invariant limit, $1/a_s \to 0$, the equation of motion Eq.~(\ref{B:eom_1}) gives the conformal solution, Eq.~(\ref{lambda}).

From several explicit calculations of the change in the pressure and the bulk viscosity near resonance \cite{Schaefer13, Nishida19, Enss19, Hofmann20}, and the scaling ansatz in Eq.~(\ref{B:scaling_ansatz}), it is possible to show that the time-dependent bulk viscosity has the approximate scaling form:

\begin{align}
\int \frac{d{\bf r}}{N} \Delta P(t) &\approx \frac{1}{\lambda(t)} \int \frac{d{\bf r}}{N}\Delta P(0) \nonumber \\
\int \frac{d{\bf r}}{N} \zeta({\bf r},t) &\approx \lambda^2(t)\int \frac{d{\bf r}}{N} \zeta({\bf r},0).
\end{align} 

\noindent This leads us to the final equation for the moment of inertia:

\begin{align}
\frac{d^2 \lambda^2(t)}{dt^2} &\approx  2(\omega_0^2 + \omega_f^2) - 4\omega_f^2 \lambda^2(t) \nonumber \\
&+ \frac{6}{m \langle r^2 \rangle(0)} \int \frac{d{\bf r}}{N} \Delta P(0)\left(\frac{1}{\lambda(t)} - 1 \right) \nonumber \\ 
&- \frac{9}{m\langle r^2 \rangle(0)} \int \frac{d {\bf r}}{N} \zeta({\bf r},0) \frac{d\lambda^2(t)}{dt}
\end{align}

\noindent which is equivalent to Eq.~(\ref{eq:hydrodynamics}). 

\section{Heisenberg Equation of Motion for the Moment of Inertia}
\label{app:HEOM}

In this appendix we show the solution to the Heisenberg equation of motion for the moment of inertia for the case of a scale invariant gas placed inside a harmonic potential of frequency $\omega_f$.

The motion is governed by the Hamiltonian, $H_s + \omega_f^2 C$, where the definitions of, $C$, and the scale invariant Hamiltonian, $H_s$, are shown in Eq.~(\ref{eq:def_so21}). We note that $C$ is related to the moment of inertia, $\langle r^2 \rangle$, by a constant prefactor: $\langle C \rangle = \langle r^2 \rangle/2$. For simplicity we will refer to $C$ in this appendix as the moment of inertia. From the SO(2,1), Eq.~(\ref{algebra}), one can obtain the following differential equation for, $C$:

\begin{equation}
\frac{d^2\langle C \rangle (t)}{dt^2} + 4 \omega_f^2 \langle C\rangle(t) = 2\langle H_s + \omega_f^2 C\rangle(0).
\label{heom_c}
\end{equation}

\noindent In obtaining Eq.~(\ref{heom_c}) we have used the fact that energy is conserved. 

Assuming the gas is initially stationary, i.e. $d\langle C\rangle(t)/dt = 0$, the solution to Eq.~(\ref{heom_c}) is:

\begin{align}
\langle C \rangle(t) &= \frac{\langle H_s + \omega_f^2 C\rangle(0)}{2\omega_f^2}  \nonumber \\
&+ \left(\langle C \rangle(0) - \frac{\langle H_s + \omega_f^2 C\rangle(0)}{2\omega_f^2}\right) \cos(2 \omega_f t)
\end{align}

\noindent As one can see the solution exhibits undamped oscillations at exactly $2\omega_f$. Similarly, the initial conditions only set the total energy, which is then conserved. As a result the reversible motion of the moment of inertia is independent of the initial conditions. This conclusion is consistent with the conformal symmetry arguments, and the hydrodynamic equations of motion presented in Appendix~\ref{app:hydro}.


\begin{thebibliography}{References}
\bibitem{Energy}
$T(t)$ here only carries the statistical meaning.
The total energy is $\lambda^2(t)Tr [ H_{proj}(t) \rho(t)]/2$ and is effectively measured by $\lambda^2(t) T(t) = T_0$, a 
conserved quantity.
\bibitem{current} From Fig.~(\ref{fig:entropy}), one can see that there are moments when $\omega (t) = \omega_f$, i.e. $H_{phys} = H_{proj}(t)$. However, at those times the gauge term ignored in Eq.~(\ref{DM}) is non zero, i.e. there is a net hydrodynamic flow. The physical state is related to the equilibrium state of the projective Hamiltonian by a gauge transformation.
\bibitem{recurrence} Note that Poincare recurrence time in a generic many-body system is practically infinite, as it is inversely proportional to the density of states, which is exponentially small at high energies. This time scale is also much larger than the Poincare recurrence time for one-particle quantum mechanics in a harmonic potential. The emergent many-body revival in the conformal expansion suggested by both the density matrix dynamics, Eq.~(\ref{DM}), and hydrodynamics, Eq.~(\ref{eq:hydrodynamics}), is not associated with Poincare recurrence in the one-particle motion, but is a pure many-body phenomenon that only occurs in the presence of the dynamic conformal symmetry.
\end{thebibliography}
\end{document}